# Improved magnetization in sputtered dysprosium thin films

G. Scheunert[1]*, W. R. Hendren[1], A. A. Lapicki[2], R. Hardeman[2], M. Gubbins[2], and R. M. Bowman[1#]

[1]Centre for Nanostructured Media, School of Mathematics and Physics, Queen's University Belfast, Belfast BT7 1NN, UK

[2]Seagate Technology, 1 Disc Drive, Springtown, Derry BT48 0BF, UK

**Abstract**

50nm thick nanogranular polycrystalline dysprosium thin films have been prepared via ultra-high vacuum DC sputtering on $SiO_2$ and Si wafers. The maximum in-plane spontaneous magnetization at T = 4K was found to be $\mu_0 M_{S,4K}^{(C)} = (3.28\pm0.26)$T for samples deposited on wafers heated to 350°C with a Neel point of $T_N^{(C)} = (173\pm2)$K and a ferromagnetic transition at $T_C^{(C)} = (80\pm2)$K, measured via zero-field-cooled – field-cooled magnetization measurements, close to single-crystal values. The slightly reduced magnetization is explained in the light of a metastable face-centered cubic crystal phase which occurred at the seed interface and granularity related effects, that are still noticeably influential despite an in-plane magnetic easy axis. As deposited samples showed reduced magnetization of $\mu_0 M_{S,4K}^{(A)} = (2.26\pm0.18)$T, however their ferromagnetic transition shifted to a much higher temperature of $T_C^{(A)} = (172\pm2)$K and the antiferromagnetic phase was completely suppressed probably as a result of strain.

Corresponding authors: * gscheunert01@qub.ac.uk    # r.m.bowman@qub.ac.uk

**Introduction**

With its outstandingly high saturation magnetization close to T = 0K of $\mu_0 M_{S,0K}$ = 3.7T in single-crystals with a ferromagnetic-antiferromagnetic (FM-AFM) metamagnetic transition at $T_C$ = 85K and a Neel point of $T_N$ = 179K [1] [2] Dy quickly was considered a possible solenoid pole piece material for ultra-high moment applications at cryogenic temperatures [3] [4] [5]. Attempts to move away from expensive single-crystals towards cheaper polycrystalline compounds lead to saturation values at T = 4.2K of still up to $\mu_0 M_{S,4.2K}$ = 3.5T for magnetically aligned Dy [6]. Similarly textured bulk Dy was recently proposed as pole material for field concentrators [7] and undulators [8] [9] whereas some other recent applications such as tips for magnetic resonant force microscopy [10] require nanometer dimensions which can best be served by thin-film deposition techniques like molecular beam epitaxy, pulsed-laser deposition or plasma sputtering, as used for this study for its high yield and scalability. Due to the large magnetic anisotropy of Dy and its interesting coupling behavior with transition metals and other ferromagnetic rare earths, often explained by Rudermann-Kittel-Kasuya-Yosida (RKKY) interactions, much attention has also been paid to Dy/$X_1$ ($X_1$ = Fe, Co, Ni) [11] [12] [13] and Dy/$X_2$ ($X_2$ = Er, Ho, Gd) [14] [15] [16] [17] [18] [19] superlattices, DyFeCo alloys in magnetic recording media [20], FeNdDyB permanent magnets with immensely high anisotropy and improved heat resistance over standard FeNdB [21], and ultra-high low-temperature coercivity FeDyTb alloys [22]. Lattice mismatch of the seed layers and resulting structural transformation even for thinnest Dy films might be the primary reason why promising parallel-coupled Dy and Fe thin films in Dy/Cr/Fe stacks [23] have not been realized to date. Coming back to initial high-moment application ideas such as pole pieces but in thin-film form [24], with the additional benefit of reduced stray fields due to its large anisotropy, in this study we present results for the deposition of high-purity Dy films on $SiO_2$ and Si wafers at room and elevated temperatures that lead to an in-plane high-field spontaneous magnetization of up to $\mu_0 M_{S,4K}^{(C)}$ = (3.28±0.26)T, unchallenged by any previous report on sputtered Dy thin films.

**Fabrication**

Films of 50nm Dy sandwiched by 5nm Ta seed and cap layers were prepared by direct-current (DC) magnetron-supported sputtering on Si & Si/300nm thermal $SiO_2$ wafers under ultra-high vacuum (UHV) conditions (base pressure $< 3 \times 10^{-9}$ mbar; argon process gas pressure $= 4 \times 10^{-3}$ mbar) using a 99.8% pure Dy target. The sputtering targets were placed a distance of 16cm under an angle of 15° with respect to the substrate in a Kurt J. Lesker co-sputtering system. For the sake of better comparability we give values of deposition rates measured by a quarz crystal at a distance of 1cm to the wafer with thicknesses additionally confirmed by X-ray diffraction (XRD). To aid uniformity the wafer was rotated at 20RPM. Oxygen contamination was minimized by pre-sputter of the targets (30min at 15 W/in$^2$) and baking the sample holder and substrate for 30min at ~350°C to allow it to outgas completely before deposition. Ta was used as a capping material to prevent further corrosion under standard atmospheric conditions as well as oxygen from the substrate's oxide layer to diffuse into the Dy [25,26].

Three distinctively different Dy thin-film types with a thickness of 50nm were achieved for different fabrication parameters as follows: Type (A) was deposited at room temperature and at a deposition rate of 0.3 – 4.6 Å/s; Type (B) was deposited on a heated substrate (350°C) at low deposition rates (~0.3 Å/s); and Type (C) was deposited on a heated substrate (350°C) at higher deposition rates (1.4 – 4.6 Å/s). A fourth reference layer of 5nm thickness, Type (R), was deposited at a deposition rate of 1.4Å/s whereas the substrate temperature did not impact on the resulting crystal structure. Henceforth, the different film types are designated by superscripts.

**Structure**

A crystallographic analysis revealed a polycrystalline-nanogranular structure of the Dy, mainly concluded from Φ-angle independent XRD patterns with significant peak broadening. The stacks of Ta(5nm)/Dy(50nm)/Ta(5nm) on Si and SiO$_2$ wafers have a clean interface at the seed layer and a rough interface at the cap which is due to increased grain formation towards the top, consistent with Stranski-Krastanov growth, similar to previous reports on vapor-deposited Dy on W crystals [27]. This can be derived from an increased peak broadening in the XRD spectra for grazing incidence scans, which primarily give information on the surface structure, compared to Θ-2Θ scans, as shown in figure 1 (a) and (b), respectively, taken at Φ = 45° to reduce the contribution of the Si substrate.

For all samples the majority of the grains had hexagonal close-packed (hcp) structure with the c-axis pointing out of the wafer plane, evident from dominating (0 0 2) peaks in the Θ-2Θ scans and comparatively small (1 0 0) peaks that were discernible on closer examination. A smaller fraction of a metastable face-centered cubic (fcc) phase was found predominately for Type (A) samples. For these samples, deposited at room temperature, the hcp grain lattice parameters were slightly strained with a = (3.64±0.02)Å and c = (5.70±0.02)Å, and the fcc lattice parameter was a = (5.21±0.02)Å, similar to what has been reported before for Dy thin films vapor deposited on carbon coated glass slides [28]. Type (B) & (C) samples had similar hcp lattice parameters of a = (3.57±0.02)Å and c = (5.65±0.02)Å which correspond closely to single-crystal values [29]. Although the hcp (1 0 0) peak was not visible in grazing incidence scans of Type (B) & (C) layers, the hcp (0 0 2) peak indicates increasing strain towards the surface as it is shifted towards smaller angles resulting in changed lattice parameters c = (5.71±0.02)Å, reminiscent of Type (A) Θ-2Θ values. This feature is not observed for measurements at Φ = 0, i.e. there is a preferred in-plane orientation for the hcp grains forming towards the surface. No difference was found between samples prepared on Si or SiO$_2$ wafers. In Type (B) & (C) layers the fcc phase was present in much smaller quantities compared to Type (A) samples with the same lattice parameter of a = (5.21±0.02)Å. In none of the samples did significant traces of the fcc (1 1 1) peak occur in grazing incidence scans, i.e. the fcc phase primarily exists at the seed interface and disappears with further nucleation towards the surface. To confirm this structural model, Ta(5nm)/Dy(5nm)/Ta(5nm) references were sputtered on Si wafers, Type (R), which did not show any traces of the hcp phase in Θ-2Θ and grazing incidence scans (figure 1 orange branches) but only the fcc (1 1 1) peak. Scherrer analysis of the thicker layers' hcp (0 0 2) peaks revealed an average grain size of ~20nm for Type (A) films and ~30nm for Type (B) & (C) layers. Note, only a generic background correction was applied, therefore the absolute grain diameters have a considerable error margin, which does not impact on relative size relations. Comparing the relative intensity of the hcp (0 0 2) peak to the substrate's Si (0 0 2) reveals a major structural difference between the two types of heated depositions: Type (C) is much better textured than Type (B). Type (A) films are considerably more poorly textured as their hcp (0 0 2) peak does not exceed the Si (0 0 2) peak's intensity in Θ-2Θ scans.

As the Dy thin films of this study show remarkable structural similarities to previous reports on Gd thin films [30] it is appropriate to draw comparisons. In both cases the fcc content was particularly pronounced for room temperature depositions close to the seed layer. In the case of Gd, stacking faults were identified as the main reason for deviations from the hcp at the seed [31], which might also apply in similar ways to Dy thin films, where cubic Ta promotes fcc growth at the seed interface. Thicker Dy layers then grow in hcp grains with the c-axis pointing out-of-plane. This is in full agreement with early observations by A. E. Curzon and H. G. Chlebek who reported fcc growth for thinner layers of Gd and Dy sandwiched by carbon, which change to a hcp structure for thicker layers [28]. As in the case of Gd, samples sputtered at elevated substrate temperatures of 350°C [32], Types (B) & (C), have a drastically reduced overall fcc content. As there are no structural differences between Type (R) layers deposited at room temperature and 350°C, heating the substrate likely only reduces the fcc bottomlayer thickness beyond 5nm. Therefore, the lattice parameters of the fcc phase are identical for all layer types, whereas the hcp ones differ. Further similarities between Gd and Dy are very similar lattice parameters of the hcp phase for heated depositions, Types (B) & (C), with single-crystal values, as well as a distortion for room temperature depositions, Type (A), which resulted in stressed hcp lattice parameters. The Ta seed layer did provide an imperfect structural dictate for the Dy hcp phase resulting in inelastic strain for thicker Type (A) layers, reminiscent to what was reported for Lu/Dy multilayers [33,34], and a change in the orientation of the grains: the (0 0 2) peak for Type (A) nearly vanished towards the surface revealing what appears to be the hcp (1 0 1) peak [35,36] in the grazing incidence scan.

**Magnetics**

Magnetic measurements parallel to the in-plane direction of the films were conducted with a Quantum Design MPMS XL SQUID magnetometer, which provided magnetic fields of up to $\mu_0 H = 5T$ and reliable measurements of magnetic moment values with an uncertainty of ~5% for values as small as $10^{-6}$ emu. Due to the small Dy volume Type (R) measurements are prone to background noise and magnetic values are given without uncertainties. Wafers were cut into pieces of ~5x5mm$^2$ and the exact surface was determined via digital imaging. There is limited literature on spontaneous magnetization in high fields of thin films of Dy available and those indicate a reduced moment [33,37,38,39,40,41] compared to bulk samples, possibly due to the microstructure, coupling to the surrounding layers, interface effects and the transition from 3D to an approximate 2D magnetic system. To establish an understanding of the saturation behavior, in-plane measurements of the spontaneous magnetization at $\mu_0 H = 5T$ were carried out. Similar to other ferromagnetic rare earths, granular Dy is not fully saturated at $\mu_0 H = 5T$ [42], in fact fields as high as $\mu_0 H = 50T$ are necessary to fully saturate polycrystalline Dy due to the high anisotropy of grains [6]. Hence, in this study we use the terminology "high-field spontaneous magnetization" or "maximal magnetization" for measurements in fields of $\mu_0 H = 5T$ despite the broad use of "saturation magnetization" for similar high-field measurements in literature. Low-field magnetization measurements at $\mu_0 H = 0.01T$, i.e. zero-field cooled – field-cooled (ZFC-FC) measurements, were preceded by a 30min bake at $T = 320°C$ to assure full demagnetization of the specimen.

In agreement to $\Phi$-angle independent $\Theta$-$2\Theta$ patterns, magnetization values are similar for all in-plane directions due to a random in-plane orientation of the nanoscale Dy grains. Measurements of the spontaneous magnetization in high fields are shown in figure 2 (a): Type (C) samples of 50nm thickness exhibit highest values of $\mu_0 M_{S,4K}^{(C)} = (3.28\pm0.26)T$ at $T = 4K$. These were the films with the best texture, largest grain size, single-crystal like hcp lattice, and lowest fcc phase content. Magnetization measurements of Type (R) films, i.e. fcc phase Dy, indicate strongly reduced maximal magnetization well below $\mu_0 M_{S,4K}^{(R)} \sim 1T$. As the fcc phase primarily occurs at the seed interface, thicker layers have a smaller relative content and, hence, they have higher high-field spontaneous magnetization as shown for a 86nm layer with $\mu_0 M_{S,4K}^{(C)} = (3.35\pm0.27)T$. Type (B) show drastically reduced values of $\mu_0 M_{S,4K}^{(B)} = (2.34\pm0.19)T$, probably due to the texture that is reduced relative to Type (C). For Type (A) films $\mu_0 M_{S,4K}^{(A)} = (2.26\pm0.18)T$, similar to Type (B), likely as a result of their comparatively high fcc phase content, strained hcp lattice, and poorest texture. A small increase in magnetization was noticeable for higher deposition rates, which positively impacts on the texture as in the case of Type (B) & (C) films. Considering the high anisotropy of Dy [43][44], the smaller average crystal size of Type (A) films, resulting in more in-plane randomly aligned grains, might have a reducing influence on the maximum magnetization, too.

The ferromagnetic states of all sample types were investigated in-depth by recording in-plane hysteresis loops at $T = 4K$: those for Type (A), (B) & (C) are shown in figure 2 (b). Measured values for coercivity were $\mu_0H_{C,4K}^{(A)} = (1.10\pm0.06)$T, $\mu_0H_{C,4K}^{(B)} = (0.18\pm0.01)$T, $\mu_0H_{C,4K}^{(C)} = (0.21\pm0.01)$T, $\mu_0H_{C,4K}^{(R)} \sim 0.075$T and remnant fields were $\mu_0M_{rem,4K}^{(A)} = 1.18$T, $\mu_0M_{rem,4K}^{(B)} = 1.50$T, $\mu_0M_{rem,4K}^{(C)} = 2.31$T, $\mu_0M_{rem,4K}^{(R)} \sim 0.15$T for Type (A), (B), (C) & (R) films, respectively. As low-field measurements suffer from an increased influence of sample size and geometry, uncertainties for remnant fields are not given. Their trend is still in line with expectations, where Type (C) layers with best texture and large grains have strongest remnant fields, whereas Type (A) films with smallest grains, high fcc content, and worst hcp texture are the opposite. Type (R) layers have smallest remnant fields in accordance with them having smallest maximum magnetization. The coercive field strength scales with grain size, but the significant difference between room temperature and heated depositions is rather surprising. Type (B) & (C) films are essentially similar and the slightly smaller Type (B) coercivities are in agreement with slightly bigger grain diameters. The much smaller grain size of Type (A) films are likely the main reason for five times larger coercive fields compared to Types (B) & (C) [45]. There might also be a contribution of intergrain anisotropies which provides an additional obstacle to switching the magnetization of the grains. Comparably high coercivity values at temperatures close to T = 0K of $H_C \sim 1$T were reported before e.g. for Zr/Dy(3nm)/Zr [46], Y/Dy(60nm)/Y [14] and $Y_{0.45}Lu_{0.55}$/Dy(5nm)/ $Y_{0.45}Lu_{0.55}$ [34] where they were linked to basal plane strain, observed earlier in Type (A) samples. The remarkably small coercivities of Type (R) films are an indication for epitaxial fcc films lacking grain boundaries.

The rather complicated magnetic transition behavior was investigated by ZFC-FC measurements. Figure 3 (a) shows the results for depositions on a heated wafer. In general, Type (B) & (C) films show in-plane magnetization behavior, which is reminiscent of the basal plane, i.e. the easy axis, in a single crystal [2,47,48,49]. From the ZFC branches it can be seen that the larger grain sizes of those films lead to $T_C$ and $T_N$ points close to but slightly below single-crystal values of $T_C^{(B)} = (76\pm2)$K, $T_N^{(B)} = (171\pm2)$K; $T_C^{(C)} = (80\pm2)$K and $T_N^{(C)} = (173\pm2)$K for Type (B) & (C) films, respectively.

Slightly higher $T_C$ and $T_N$ values for Type (C) samples are likely the result of better texture and minimized residual stress. To confirm this two Type (C) samples were post-annealed for 2 hours at 400°C (pressure ~ $10^{-6}$ mbar) in a magnetic field of $\mu_0 H = 2T$ applied in-plane which indeed lead to increased transition temperatures of $T_C = (83\pm2)$K and $T_N = (174\pm2)$K. Corresponding FC branches of Type (B) & (C) layers show a bifurcation at $T_N$ and an even smoother AFM-FM transition, where magnetization values are lower than those of the ZFC branch. This somewhat surprising behavior was also observed for Dy-Y superlattices [37] and might be due to thermal hysteresis arising from coexisting helimagnetic states of opposite chirality [50].

Figure 3 (b) shows measurements for room temperature depositions and the fcc reference. Type (R) ZFC-FC branches bifurcate at $T_B^{(R)} = (115\pm2)$K and there is a Curie temperature at $T_C^{(R)} = (155\pm2)$K. There is no evidence of an AFM phase and thus it is expected that the film is FM throughout. Type (A) films have ZFC-FC characteristics significantly different to those of Types (B) & (C), exhibiting multiple magnetic transitions. Magnetization drops to zero at $T_C^{(A)} = (172\pm2)$K, followed by a bifurcation of the ZFC-FC curves at $T_B^{(A)} = (165\pm2)$K, and another two transitions at $T_C = (155\pm2)$K and $T_B = (115\pm2)$K, which match Type (R) Curie and bifurcation temperatures, respectively. As Type (A) films are comprised of an fcc layer at the seed and strained hcp grains on top, its magnetics can be understood as a superposition of both. Hence, $T_C^{(A)}$ and $T_B^{(A)}$ can be attributed to the strained hcp phase and the transitions at 155K and 115K are manifestations of the fcc phase, which has properties similar to Type (R) films. The absence of those fcc-related transitions in Type (B) & (C) films is due to a much smaller fcc content, but also similar coercivities: As $H_C^{(R)}$ is much smaller than $H_C^{(A)}$, the magnetization signal of Type (R) becomes disproportionally large in Type (A) layers, particularly for low-field measurements.

There is no FM-AFM transition in Type (A) films, the FM state persisting up to the Curie temperature at $T_C^{(A)} = (172\pm2)$K. This is likely a result of epitaxial strain at the seed interface: seed materials with a lattice mismatch, specifically with smaller hcp lattice parameters such as Er [14][15][16], Lu [33][51], and Lu-rich $Y_XLu_{1-X}$ alloys [34], have been shown to lead to positive epitaxial strain and increased transition temperatures [52] and, in the case of Lu, also to a completely suppressed AFM phase. It is worth noticing that similar differences in the ZFC-FC magnetization behavior, as shown here for Type (A) to (C) films, were reported before for MBE-grown Dy on Zr seed layers [40][46], where thicker Dy films exhibit magnetics akin to our Type (B) & (C) whereas thinner Dy layers are similar to Type (A) films, which suggests a change in crystal structure with further nucleation as for Dy films in this study.

**Summary & Conclusion**

The properties of magnetron sputtered Dy films deposited at room temperature and 350°C on Ta seed layers on $SiO_2$ and Si wafers were studied by using XRD and magnetic ZFC-FC SQUID measurements. Depostions at 350°C had hcp (0 0 1) texture, and magnetic properties very close to Dy single-crystals, such as a FM-AFM transition and Neel point of $T_C^{(C)} = (80\pm2)$K and $T_N^{(C)} = (173\pm2)$K, respectively, and a high-field spontaneous magnetization at T = 4K of $\mu_0 M_{S,4K}^{(C)} = (3.28\pm0.26)$T. By contrast, depositions at room temperature had higher granularity and a crystal structure that changed from fcc at the seed layer interface to hcp for the bulk of the film. Magnetic properties showed lower maximum magnetization and higher coercivity at T = 4K, and no FM-AFM transition, the films being ferromagnetic up to a Curie temperature of $T_C^{(A)} = (172\pm2)$K. The properties of the sputtered films were comparable to similar studies on films grown epitaxially by MBE.

**Figure Captions:**

FIGURE 1: XRD spectra of $\Theta$-$2\Theta$ scans, graph (a), and grazing incidence scans, graph (b), of Ta(5nm)/Dy(50nm)/Ta(5nm) trilayers sputtered on Si wafers, branches (A), (B) & (C) correspond to the according film types. The orange branch (R) shows a Ta(5nm)/Dy(5nm)/Ta(5nm) reference layer. Also visible for all samples is the Si (0 0 2) peak at $2\Theta = 33°$. Note, the substrate was rotated $\Phi = 45°$ for the grazing incidence scan to reduce background contribution of the wafer.

FIGURE 2: In-plane high-field ($\mu_0 H = 5T$) spontaneous magnetization of pure Dy layers on Si & $SiO_2$ wafers with respect to layer thicknesses, graph (a). Hysteresis loops recorded in-plane at T = 4K, graph (b), the maximum applied field was $\mu_0|H_{max}| = 5T$ shown in the inset.

FIGURE 3: In-plane ZFC($\rightarrow$)–FC($\leftarrow$) magnetization measurements ($\mu_0 H = 0.01T$) indicating magnetic transitions such as $T_N$ and $T_C$. Due to shape anisotropy sample size and geometry massively impact on the absolute magnetization values.

FIGURE 1 (a)  FIGURE 1 (b)

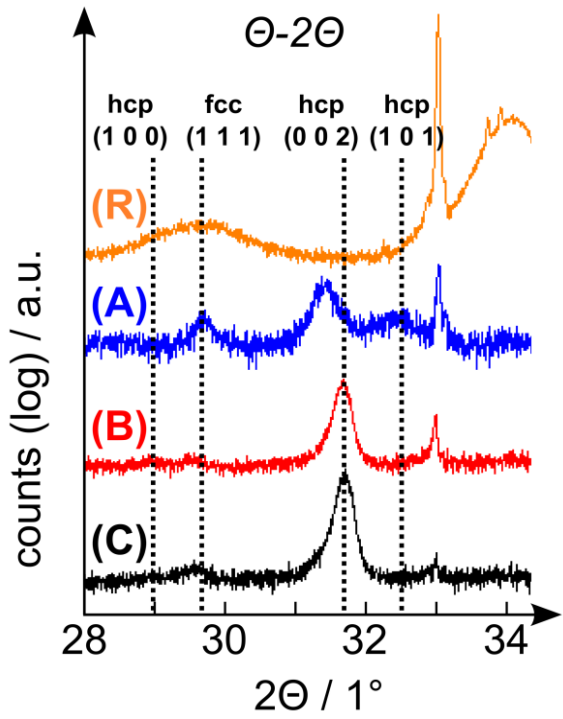
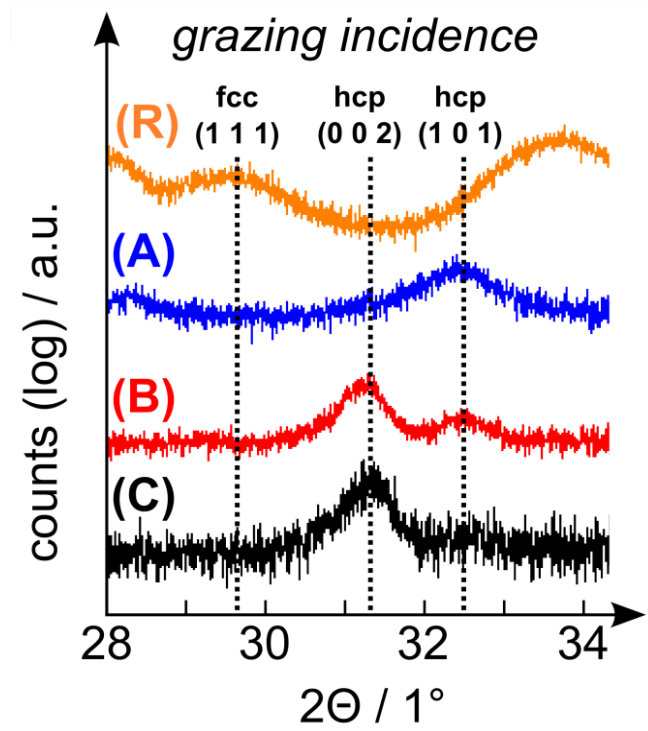

FIGURE 2 (a): 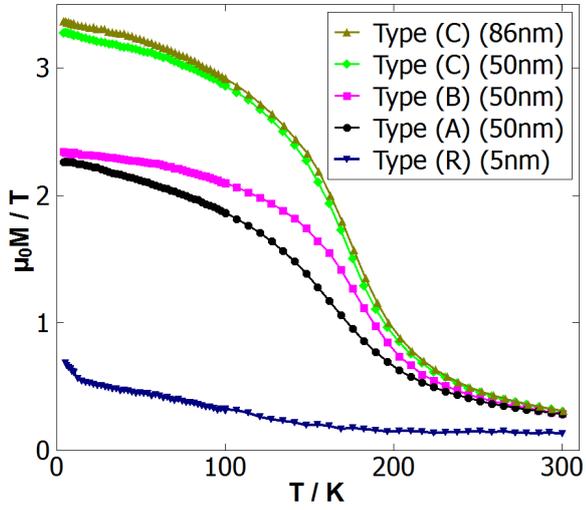

FIGURE 2 (b): 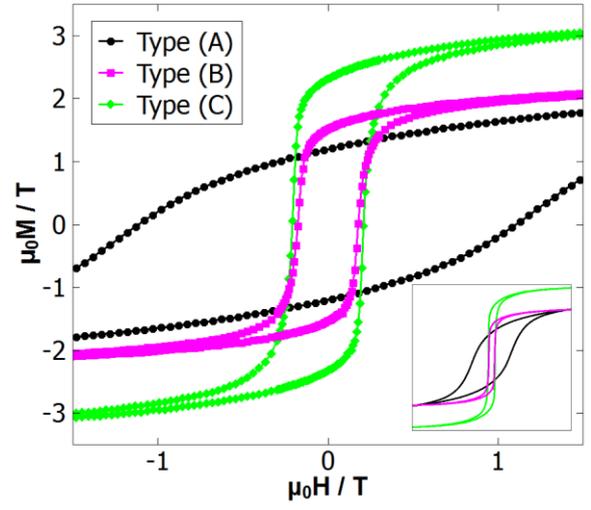

FIGURE 3 (a): 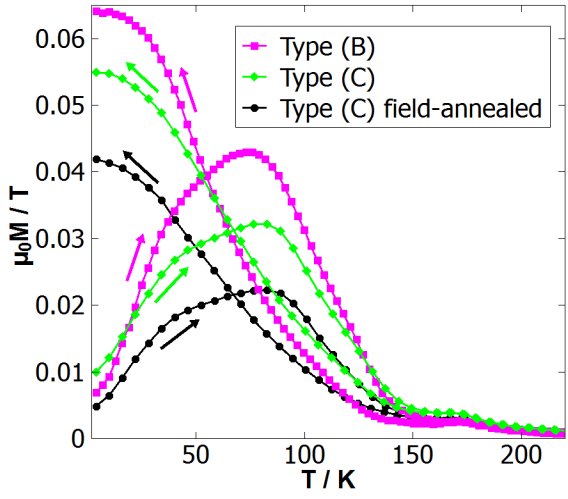

FIGURE 3 (b): 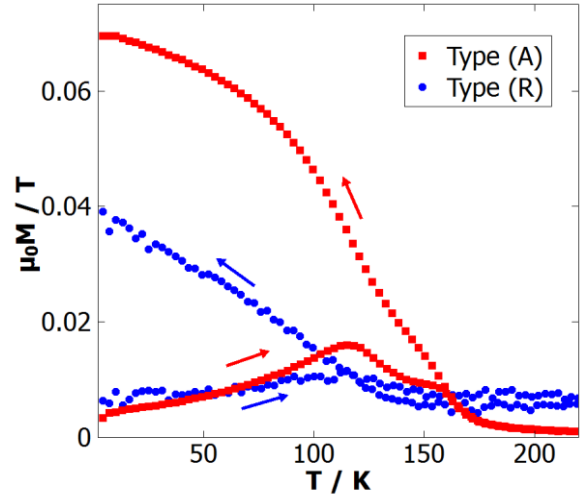